\documentclass{llncs}

\usepackage[usenames]{color}

\usepackage{latexsym}
\usepackage{amsfonts}
\usepackage{graphicx}
\usepackage{amsmath}
\usepackage{amssymb}

\newcommand{\Z}{\mathbb{Z}}

\begin{document}

\title{Using semidirect product\\
 of (semi)groups in public key cryptography}
\author{Delaram Kahrobaei\inst{1}  \and Vladimir Shpilrain\inst{2}}
\institute{CUNY Graduate Center and City Tech, City University of
New York \email{dkahrobaei@gc.cuny.edu} \thanks{Research of Delaram
Kahrobaei was partially supported by a PSC-CUNY grant from the CUNY
research foundation, as well as the City Tech foundation. Research
of Delaram Kahrobaei and Vladimir Shpilrain was also supported by
the ONR (Office of Naval Research) grant N000141512164.} \and The
City College of New York and CUNY Graduate Center
\email{shpil@groups.sci.ccny.cuny.edu}
\thanks{Research of Vladimir Shpilrain was partially supported by the NSF grant CNS-1117675.}}

\maketitle

\begin{abstract}
In this survey, we   describe a general key exchange protocol based
on  semidirect product of (semi)groups (more specifically, on
extensions of (semi)groups by automorphisms), and then focus on
practical instances of this general idea. This protocol can be based
on any group or semigroup, in particular on any non-commutative
group. One of its special cases is the standard Diffie-Hellman
protocol, which is based on a cyclic group. However, when this
protocol is used with a non-commutative (semi)group, it acquires
several useful features that make it compare favorably to the
Diffie-Hellman protocol. The focus then shifts to selecting an
optimal platform (semi)group, in terms of security and efficiency.
We show, in particular, that one can get a variety of new security
assumptions by varying an automorphism used for a (semi)group
extension.

\end{abstract}

\section{Introduction}

The area of public key cryptography started  with the seminal paper
\cite{DH} introducing what is now known as the Diffie-Hellman key
exchange protocol.

The simplest, and original, implementation of the protocol uses the
multiplicative group of integers modulo $p$, where $p$ is prime and
$g$ is primitive $\mod p$. A more general description of the
protocol uses an arbitrary finite cyclic group.

\begin{enumerate}

\item Alice and Bob agree on a finite cyclic group $G$ and a generating element $g$ in $G$.
 We will write the group $G$ multiplicatively.

\item Alice picks a random natural number $a$ and sends $g^a$ to Bob.

\item    Bob picks a random natural number $b$ and sends $g^b$ to Alice.

\item   Alice computes $K_A=(g^b)^a=g^{ba}$.

\item  Bob computes $K_B=(g^a)^b=g^{ab}$.
\end{enumerate}

Since $ab=ba$, both Alice and Bob are now in possession of the same
group element $K=K_A= K_B$ which can serve as the shared secret key.

The protocol is considered secure against eavesdroppers if $G$ and
$g$ are chosen properly. The eavesdropper must solve the {\it
Diffie-Hellman problem} (recover $g^{ab}$ from $g$, $g^a$ and $g^b$)
to obtain the shared secret key. This is currently considered
difficult for a ``good" choice of parameters (see e.g.
\cite{Menezes} for details).

There is  an ongoing search for other platforms where the
Diffie-Hellman or similar key exchange could be carried out more
efficiently or where security would be based on different
assumptions. This search already gave rise to several interesting
directions, including a whole area of elliptic curve cryptography
\cite{Washington}. We also refer the reader to \cite{MSU} or
\cite{MSU2} for a survey of proposed cryptographic primitives based
on non-abelian (= non-commutative) groups. A survey of these efforts
is outside of the scope of the present paper; our goal here is to
describe a new  key exchange protocol from \cite{HKS}  based on
extension of a (semi)group by automorphisms (or more generally, by
self-homomorphisms) and discuss possible platforms that would make
this protocol secure and efficient. This protocol can be based on
any group, in particular on any non-commutative group. It has some
resemblance to the classical Diffie-Hellman protocol, but there are
several distinctive features that, we believe, give the new protocol
important advantages. In particular, even though the parties do
compute a large power of a public element (as in the classical
Diffie-Hellman protocol), they do not transmit the whole result, but
rather just part of it.

We then describe in this survey some particular instantiations of
this general protocol.  We start with a non-commutative semigroup of
matrices as the platform, consider an extension of this semigroup by
a conjugating automorphism and show that security of the relevant
instantiation is based on a quite different security assumption
compared to that of the standard Diffie-Hellman protocol. However,
due to the nature of this security assumption, the protocol turns
out to be vulnerable to a ``linear algebra attack", similar to an
attack on Stickel's protocol \cite{Stickel_self} offered in
\cite{Stickel}, albeit more sophisticated, see \cite{MR}, \cite{R}.
A composition of conjugating automorphism with a field automorphism
was employed in \cite{KLS}, but this automorphism still turned out
to be not complex enough to make the protocol withstand a linear
algebra attack, see \cite{DMU}, \cite{R}.

We therefore offer here another platform group that we believe
should make the protocol invulnerable to the attacks of \cite{DMU},
\cite{MR}, \cite{R}. The group is a {\it free nilpotent $p$-group},
for a sufficiently large prime $p$. We give a formal definition of
this group in Section \ref{nilpotent}; here we just say that this is
a finite group all of whose elements have order dividing $p^n$ for
some fixed $n \ge 1$. As any finite group, this group is linear, but
Janusz \cite{Janusz} showed that a faithful representation of a
finite $p$-group, with at least one element of order $p^n$, as a
group of matrices over a finite field of characteristic $p$ is of
dimension at least  $1+p^{n-1}$, which is too large to launch a
linear algebra attack provided $p$ itself is large enough. At the
same time, to keep computation in the platform group efficient, the
nilpotency class of the group has to be fairly small.
We note that, in contrast, the dimension of the classical
representations of finitely generated {\it torsion-free} nilpotent
groups in a matrix group $UT(\Z)$ can be rather small (cf.
\cite{WN}), but for torsion groups with elements of large order the
situation is really different. Still, there is the usual trade-off
between security and efficiency, so the following parameters have to
be chosen carefully to provide for both security and efficiency: (1)
the size of $p$; (2) the nilpotency class of the platform group; (3)
the rank (i.e., the number of generators) of the platform group. We
discuss this in our Section \ref{nilpotent}.


We mention here another, rather different, proposal \cite{PHKCP} of
a cryptosystem based on the semidirect product of two groups and yet
another, more complex, proposal of a key agreement based on the
semidirect product of two monoids \cite{AAGL}. Both these proposals
are very different from that of \cite{HKS}.  In particular, the
crucial idea of transmitting just part of the result of an
exponentiation appears only in \cite{HKS}.

Finally, we note that the basic construction (semidirect product)
described in this survey can be adopted, with some simple
modifications, in other algebraic systems, e.g. associative rings or
Lie rings, and key exchange protocols similar to ours can be built
on those.

\section{Semidirect products and extensions  by automorphisms}
\label{Semidirect}

We include this section to make the exposition more comprehensive.
The reader who is uncomfortable with group-theoretic constructions
can skip to subsection \ref{holomorph}.

We now recall the definition of a semidirect product:

\begin{definition} Let $G, H$ be two groups, let $Aut(G)$ be the group of automorphisms of $G$,
and let $\rho: H \rightarrow Aut(G)$ be a homomorphism. Then the
semidirect product of $G$ and $H$ is the set
$$\Gamma = G \rtimes_{\rho} H = \left \{ (g, h): g \in G, ~h \in H \right \}$$
with the group operation given by\\
\centerline{$(g, h)(g', h')=(g^{\rho(h')} \cdot  g', ~h \cdot h')$.}\\
Here $g^{\rho(h')}$ denotes the image of  $g$ under the automorphism
$\rho(h')$, and when we write a product $h \cdot h'$ of two
morphisms, this means that $h$ is applied first.
\end{definition}

In this paper, we focus on a special case of this construction,
where the group $H$ is just a subgroup of the group $Aut(G)$. If
$H=Aut(G)$, then the corresponding semidirect product is called the
{\it holomorph} of the group $G$. We give some more details about
the holomorph in our Section \ref{holomorph}, and in Section
\ref{protocol_holomorph} we describe a key exchange protocol that
uses (as the platform) an extension of a group $G$ by a {\it cyclic}
group of automorphisms.

\subsection{Extensions  by automorphisms} \label{holomorph}

A particularly simple special case of the semidirect product
construction is where the group $H$ is just a subgroup of the group
$Aut(G)$. If $H=Aut(G)$, then the corresponding semidirect product
is called the {\it holomorph} of the group $G$. Thus,  the holomorph
of $G$, usually denoted by $Hol(G)$, is the set of all pairs $(g,
~\phi)$, where $g \in G, ~\phi \in Aut(G)$, with the group operation
given by ~$(g,~\phi)\cdot  (g',~\phi') = (\phi'(g)\cdot g',~\phi
\cdot \phi')$.

It is often more practical to use a subgroup of $Aut(G)$ in this
construction, and this is exactly what we do in Section
\ref{protocol_holomorph}, where we describe a key exchange protocol
that uses (as the platform) an extension of a group $G$ by a cyclic
group of automorphisms.

\begin{remark}
One can also use this construction if $G$ is not necessarily a
group, but just a semigroup, and/or consider endomorphisms (i.e.,
self-homomorphisms) of $G$, not necessarily automorphisms. Then the
result will be a semigroup; this is what we use in our Section
\ref{Matrices}.
\end{remark}

\section{Key exchange protocol}
\label{protocol_holomorph}

In the simplest implementation of the construction described in our
Section \ref{holomorph}, one can use just a cyclic subgroup (or a
cyclic subsemigroup) of the group $Aut(G)$ (respectively, of the
semigroup  $End(G)$ of endomorphisms) instead of the whole group of
automorphisms of $G$.

Thus, let $G$ be a (semi)group. An element $g\in G$ is chosen and
made public as well as an arbitrary automorphism $\phi\in Aut(G)$
(or an arbitrary endomorphism $\phi\in End(G)$). Bob chooses a
private $n\in \mathbb{N}$, while Alice chooses a private $m\in
\mathbb{N}$. Both Alice and Bob are going to work with elements of
the form $(g, \phi^r)$, where $g\in G, ~r\in \mathbb{N}$. Note that
two elements of this form are multiplied as follows:  ~$(g, \phi^r)
\cdot (h, \phi^s) = (\phi^s(g)\cdot h, ~\phi^{r+s})$.
\medskip

\begin{enumerate}

\item Alice computes $(g, \phi)^m = (\phi^{m-1}(g) \cdots \phi^{2}(g) \cdot \phi(g) \cdot g,
~\phi^m)$ and sends {\bf only the first component} of this pair to
Bob. Thus, she sends to Bob {\bf only} the element $a =
\phi^{m-1}(g) \cdots \phi^{2}(g) \cdot \phi(g) \cdot g$ of the
(semi)group $G$.
\medskip

\item Bob computes $(g, \phi)^n = (\phi^{n-1}(g) \cdots \phi^{2}(g) \cdot \phi(g) \cdot g,
~\phi^n)$ and sends {\bf only the first component} of this pair to
Alice. Thus, he sends to Alice {\bf only} the element $b =
\phi^{n-1}(g) \cdots \phi^{2}(g) \cdot \phi(g) \cdot g$ of the
(semi)group $G$.
\medskip

\item Alice computes $(b, x) \cdot (a, ~\phi^m) = (\phi^m(b) \cdot a,
~x  \cdot \phi^{m})$. Her key is now $K_A = \phi^m(b) \cdot a$. Note
that she does not actually ``compute" $x \cdot \phi^{m}$ because she
does not know the automorphism $x=\phi^{n}$; recall that it was not
transmitted to her. But she does not need it to compute $K_A$.
\medskip

\item Bob computes $(a, y) \cdot (b, ~\phi^n) = (\phi^n(a) \cdot b,
~y \cdot \phi^{n})$. His key is now $K_B = \phi^n(a) \cdot b$.
Again, Bob does not actually ``compute" $y \cdot \phi^{n}$ because
he does not know the automorphism $y=\phi^{m}$.
\medskip

\item Since $(b, x) \cdot (a, ~\phi^m) = (a, ~y) \cdot (b, ~\phi^n) =
(g, ~\phi)^{m+n}$, we should have $K_A = K_B = K$, the shared secret
key.

\end{enumerate}

\begin{remark}
Note that, in contrast with the ``standard" Diffie-Hellman key
exchange, correctness here is based on the equality $h^{m}\cdot
h^{n} = h^{n} \cdot h^{m} =  h^{m+n}$  rather  than on the equality
$(h^{m})^{n} = (h^{n})^{m} = h^{mn}$. In  the ``standard"
Diffie-Hellman set up, our trick would not work because, if the
shared key $K$ was just the product of two  openly transmitted
elements, then anybody, including the eavesdropper, could compute
$K$.
\end{remark}

\section{Computational cost}
\label{cost}

From the look of transmitted elements in the protocol in Section
\ref{protocol_holomorph}, it may seem that the parties have to
compute a product of $m$ (respectively, $n$) elements of the
(semi)group $G$. However, since the parties actually compute powers
of an element of $G$, they can use the ``square-and-multiply"
method, as in the standard Diffie-Hellman protocol. Then there is a
cost of applying an automorphism $\phi$ to an element of $G$, and
also of computing powers of $\phi$. These costs depend, of course,
on a specific platform (semi)group that is used with our protocol
and on a specific automorphism that is used for a (semi)group
extension. In our first, ``toy" example (Section \ref{Toy} below),
both applying an automorphism $\phi$ and computing its powers amount
to exponentiation of elements of $G$, which can be done again by the
``square-and-multiply" method. In our example in Section
\ref{Matrices}, $\phi$ is a conjugation, so applying $\phi$ amounts
to just two multiplications of elements in $G$, while computing
powers of $\phi$ amounts to exponentiation of two elements of $G$
(namely, of the conjugating element and of its inverse).

Thus, in either instantiation of our protocol considered in this
paper, the cost of computing $(g, \phi)^n$ is $O(\log n)$, just as
in the standard Diffie-Hellman protocol. Computational cost analysis
for the platform  group suggested in Section \ref{nilpotent} is
somewhat more delicate; we refer to Section \ref{p-group} for more
details.

%

\section{``Toy example": multiplicative $\mathbb{Z}_p^*$}
\label{Toy}

As one of the simplest instantiations  of our protocol, we use here
the multiplicative group $\mathbb{Z}_p^*$  as the platform group $G$
to illustrate what is going on. In selecting a prime $p$, as well as
private exponents $m, n$, one can follow the same guidelines as in
the ``standard" Diffie-Hellman.

Selecting the (public) endomorphism  $\phi$ of the group
$\mathbb{Z}_p^*$ amounts to selecting yet another integer $k$, so
that for every $h \in \mathbb{Z}_p^*$, one has $\phi(h) = h^k$. If
$k$ is  relatively prime to $p-1$, then $\phi$ is actually an
automorphism. Below we assume that $k > 1$.

Then, for an element $g \in \mathbb{Z}_p^*$, we have:
$$(g, \phi)^m =
(\phi^{m-1}(g) \cdots    \phi(g) \cdot \phi^{2}(g) \cdot g,
~\phi^m).$$

We focus on the first component of the element on the right; easy
computation shows that it is equal to $g^{k^{m-1} +\ldots +k +1} =
g^{\frac{k^{m}-1}{k-1}}$. Thus, if the adversary chooses a ``direct"
attack, by trying to recover the private exponent $m$, he will have
to solve the discrete log problem twice: first to recover
$\frac{k^{m}-1}{k-1}$ from $g^{\frac{k^{m}-1}{k-1}}$, and then to
recover $m$ from $k^{m}$. (Note that $k$ is public since $\phi$ is
public.)

On the other hand, the analog of what is called ``the Diffie-Hellman
problem" would be to recover the shared key $K =
g^{\frac{k^{m+n}-1}{k-1}}$ from the triple $(g,
~g^{\frac{k^{m}-1}{k-1}}, ~g^{\frac{k^{n}-1}{k-1}})$. Since $g$ and
$k$ are public, this is equivalent to recovering $g^{k^{m+n}}$ from
the triple $(g, ~g^{k^m},  ~g^{k^n})$, i.e., this is exactly the
standard Diffie-Hellman problem.

Thus, the bottom line of this example is that the instantiation of
our protocol where the group $G$ is $\mathbb{Z}_p^*$, is not really
different from the standard Diffie-Hellman protocol. In the next
section, we describe a more interesting instantiation, where the
(semi)group $G$ is non-commutative.

\section{Matrices over group rings and  extensions by inner automorphisms}
\label{Matrices}

Our exposition here follows \cite{HKS}. To begin with, we note that
the general protocol in Section \ref{protocol_holomorph} can be used
with {\it any} non-commutative group $G$  if $\phi$ is selected to
be a non-trivial inner automorphism, i.e., conjugation by an element
which is not in the center of $G$. Furthermore, it can be used  with
any non-commutative {\it semigroup} $G$ as well, as long as  $G$ has
some invertible elements; these can be used to produce  inner
automorphisms. A typical example of such a semigroup would be a
semigroup of matrices over some ring.

In the paper \cite{KKS1}, the authors have employed matrices over
group rings of a (small) symmetric group as platforms for the
(standard) Diffie-Hellman-like key exchange. In this section, we use
these matrix semigroups again and consider an extension of such a
semigroup by an inner automorphism to get a platform semigroup for
the general protocol in Section \ref{protocol_holomorph}.

Recall that a (semi)group ring $R[S]$ of a (semi)group  $S$ over a
commutative ring $R$ is the set of all formal sums $\sum_{g_i \in S}
r_i g_i$, where $r_i \in R$, and all but a finite number of $r_i$
are zero.

The sum of two elements in $R[G]$ is defined by $$\left(\sum_{g_i\in
S}a_ig_i\right)+\left(\sum_{g_i\in S}b_ig_i\right) = \sum_{g_i \in
S}(a_i+b_i)g_i.$$

The multiplication of two elements in $R[G]$ is defined by using
distributivity.

As we have already pointed out, if a (semi)group $G$ is
non-commutative and has non-central invertible elements, then it
always has a non-identical inner automorphism, i.e., conjugation by
an element $g \in G$ such that $g^{-1} h g \ne h$ for at least some
$h \in G$.

Now let $G$ be the semigroup of  $3 \times 3$ matrices over the
group ring $\mathbb{Z}_{7}[A_5]$, where  $A_5$ is the alternating
group on 5 elements. 
Here we use an extension of the semigroup $G$ by an inner
automorphism $\varphi_{_H}$, which is conjugation by a matrix $H \in
GL_3(\mathbb{Z}_{7}[A_5])$. Thus, for any matrix $M \in G$ and for
any integer $k \ge 1$, we have

$$\varphi_{_H}(M) = H^{-1} M H; ~\varphi^k_{_H}(M) = H^{-k} M H^k.$$

\noindent Now the general protocol from Section
\ref{protocol_holomorph} is specialized in this case as follows.

\medskip

\begin{enumerate}

\item Alice and Bob agree on  public matrices $M \in G$ and $H \in
GL_3(\mathbb{Z}_{7}[A_5])$. Alice selects a private positive integer
$m$, and Bob selects a private positive integer $n$.
\medskip

\item Alice computes $(M, \varphi_{_H})^m = (H^{-m+1} M H^{m-1} \cdots H^{-2} M H^2 \cdot H^{-1} M H \cdot M,
~\varphi_{_H}^m)$ and sends {\bf only the first component} of this
pair to Bob. Thus, she sends to Bob {\bf only} the matrix\\
$$A = H^{-m+1} M H^{m-1} \cdots H^{-2} M H^2 \cdot H^{-1} M H \cdot M
= H^{-m} (HM)^{m}.$$

\item Bob computes $(M, \varphi_{_H})^n = (H^{-n+1} M H^{n-1} \cdots H^{-2} M H^2 \cdot H^{-1} M H \cdot M,
~\varphi_{_H}^n)$ and sends {\bf only the first component} of this
pair
to Alice. Thus, he sends to Alice {\bf only} the matrix\\
$$B = H^{-n+1} M H^{n-1} \cdots H^{-2} M H^2 \cdot H^{-1} M H \cdot M
= H^{-n} (HM)^{n}.$$


\item Alice computes $(B, x) \cdot (A, ~\varphi_{_H}^m) = (\varphi_{_H}^m(B) \cdot A,
~x  \cdot \varphi_{_H}^{m})$. Her key is now $K_{Alice} =
\varphi_{_H}^m(B) \cdot A = H^{-(m+n)}(HM)^{m+n}$. Note that she
does not actually ``compute" $x \cdot \varphi_{_H}^{m}$ because she
does not know the automorphism $x=\varphi_{_H}^{n}$; recall that it
was not transmitted to her. But she does not need it to compute
$K_{Alice}$.
\medskip

\item Bob computes $(A, y) \cdot (B, ~\varphi_{_H}^n) = (\varphi_{_H}^n(A) \cdot B,
~y \cdot \varphi_{_H}^{n})$. His key is now $K_{Bob} =
\varphi_{_H}^n(A) \cdot B$. Again, Bob does not actually ``compute"
$y \cdot \varphi_{_H}^{n}$ because he does not know the automorphism
$y=\varphi_{_H}^{m}$.
\medskip

\item Since $(B, x) \cdot (A, ~\varphi_{_H}^m) = (A, ~y) \cdot (B, ~\varphi_{_H}^n) =
(M, ~\varphi_{_H})^{m+n}$, we should have $K_{Alice} = K_{Bob} = K$,
the shared secret key.

\end{enumerate}

\section{Security assumptions }
\label{Security}

In this section, we address the question of security of the protocol
described in Section \ref{Matrices}.

Recall that the shared secret key in the protocol of Section
\ref{Matrices} is

$$K = \varphi_{_H}^m(B) \cdot A = \varphi_{_H}^n(A) \cdot B = H^{-(m+n)}(HM)^{m+n}.$$

\noindent Therefore, our security assumption here is that it is
computationally hard to retrieve the key $K = H^{-(m+n)}(HM)^{m+n}$
from the quadruple\\
$(H, ~M, ~H^{-m}(HM)^{m}, ~H^{-n}(HM)^{n})$.

In particular, we have to take care that the matrices $H$ and  $HM$
do not commute because otherwise, $K$ is just a product of
$H^{-m}(HM)^{m}$  and  $H^{-n}(HM)^{n}$.

A weaker security assumption arises if an eavesdropper tries to
recover a private exponent from a transmission, i.e., to recover,
say, $m$ from $H^{-m} (HM)^{m}.$ A special case of this problem,
where $H=I$, is the ``discrete log" problem for matrices over
$\mathbb{Z}_{7}[A_5]$, namely: recover $m$ from $M$ and $M^{m}$.

As we have mentioned in the Introduction, the protocol in this
section was attacked in \cite{MR} and \cite{R} by a ``linear algebra
attack". This was possible partly because of the special ``compact"
form of the above security assumptions, and partly because the
dimension of a linear representation of the platform semigroup
happens to be small enough in this case for a linear algebra attack
to be computationally feasible. In the following Section
\ref{nilpotent}, we offer another platform that does not have these
vulnerabilities.

\section{Nilpotent groups and $p$-groups}
\label{nilpotent}

First we recall that a {\it free group} $F_r$ on $x_1, \ldots, x_r$
is the set of {\it reduced words} in the alphabet $\{x_1, \ldots,
x_r, x_1^{-1}, \ldots, x_r^{-1}\}$.  A reduced word is a  word
without subwords $x_ix_i^{-1}$ or $x_i^{-1}x_i$. The multiplication
on this set is concatenation of two words, followed by canceling out
all subwords $x_ix_i^{-1}$ and   $x_i^{-1}x_i$ until the word
becomes reduced.

It is a fact that every group that can be generated by $r$ elements
is the factor group of $F_r$ by an appropriate normal subgroup. We
are now going to define two special normal subgroups of $F_r$.

The normal subgroup $F_r^p$ is generated (as a group) by all
elements of the form $g^p, ~g \in F_r$. In the factor group
$F_r/F_r^p$ every nontrivial element therefore has order $p$ (if $p$
is a prime). More generally, if $n \ge 2$ is an arbitrary integer,
then the order of any element of $F_r/F_r^n$ divides $n$.

The other normal subgroup that we need is somewhat less
straightforward to define. Let $[a, b]$ denote $a^{-1} b^{-1} ab$.
Then, inductively, let $[y_1, \ldots, y_{c+1}]$ denote $[[y_1,
\ldots, y_{c}], y_{c+1}]$. For a group $G$, denote by $\gamma_c(G)$
the (normal) subgroup of $G$ generated (as a group) by all elements
of the form $[y_1, \ldots, y_{c}]$. If $\gamma_{c+1}(G)=\{1\}$, we
say that the group $G$ is nilpotent of nilpotency class $c$.

The factor group $F_r/\gamma_{c+1}(F_r)$ is called {\it the free
nilpotent group} of nilpotency class $c$. This group is infinite;
however, the group we define in the following subsection is finite,
and we are going to recommend it as the platform for the
cryptographic scheme based on a semidirect product.

\subsection{Free nilpotent $p$-group}
\label{p-group}

The group $G=F_r/F_r^{p^2} \cdot \gamma_{c+1}(F_r)$ is what we
suggest to use as the platform for the key exchange protocol in
Section \ref{protocol_holomorph}.

This group, being a nilpotent $p$-group, is finite. Its order
depends on $p$, $c$, and $r$. For efficiency reasons, it seems
better to keep $c$ and $r$ fairly small (in particular, we suggest
$c=2$ or $3$), while $p$ should be large enough to make the
dimension of linear representations of $G$ so large that a linear
algebra attack would be infeasible. As we have mentioned in the
Introduction, a faithful representation of a finite $p$-group, with
at least one element of order $p^n$, as a group of matrices over a
finite field of characteristic $p$ is of dimension at least
$1+p^{n-1}$ \cite{Janusz}, so in our case it is of dimension at
least  $1+p$. Thus, if $p$ is, say, a 100-bit number, a linear
algebra attack is already infeasible.

At the same time, we want computation in the group $G$ to be
efficient. Also, we want transmitted elements to be in some kind of
standard form, usually called a {\it normal form}. Here is how a
normal form looks like if nilpotency class $c=2$: $$x_1^{\alpha_1}
\cdots x_i^{\alpha_i} \cdots x_r^{\alpha_r} [x_1,x_2]^{\beta_{1,2}}
\cdots [x_i,x_j]^{\beta_{i,j}} \cdots
  [x_{r-1},x_r]^{\beta_{r-1,r}},$$
where $\alpha_i$ and  $\beta_{i,j}$ are integers and  in every
$[x_i,x_j]$ above one has $i<j$. Different collections of $\alpha_i$
and $\beta_{i,j}$ produce different elements of $G$ as long as $0
\le \alpha_i, \beta_{i,j} < p^2$, so $G$ in this case has at least
$p^{2r + r(r-1)}= p^{r^2+r}$ elements, which is a large number even
if $r$ is fairly small. At the same time, group operations (i.e.,
multiplication and inversion) in $G$ are quite efficient. Indeed,
multiplying two elements in the above form essentially amounts to
re-writing a product $x_1^{\alpha_1} \cdots x_r^{\alpha_r}  \cdot
x_1^{\alpha'_1} \cdots x_r^{\alpha'_r}$ in the normal form. This is
because commutators $[x_i,x_j]$ commute with any element of $G$
(since $c=2$), so collecting all $[x_i,x_j]$ in the right place
takes (almost) linear time in the length of an input. Now re-writing
a product of powers of $x_i$ in the normal form is not too hard
either because $[x_i^a, x_j^b]=[x_i,x_j]^{ab}$ in the group $G$
(again, since $c=2$). Thus, re-writing will take at most quadratic
time in the length of an input.

Applying an endomorphism (i.e., a self-homomorphism) $\phi$ given as
a map $\phi(x_i)=y_i$ on the generators is efficient, too. This is
due to the fact that in any group $G$ of nilpotency class 2, one
has: (1) $ab=ba$ if either $a$ or $b$ (or both) belong to
$\gamma_2(G)$; ~(2) $[ab,c]=[a,c][b,c]$ and $[a,bc]=[a,b][a,c]$ for
any $a, b, c \in G$; (3) $(ab)^n = a^n b^n [b,
a]^{\frac{n(n-1)}{2}}$ for any $a, b \in G$. Using these identities,
one can reduce $\phi(g)$ to the normal form in at most quadratic
time in the length of $g \in G$, provided $g$ itself was in the
normal form.

The group $G$ has another property useful for our purposes. We note
that the subgroup $F_r^{p^2} \cdot \gamma_{c+1}(F_r)$ of $F_r$ is,
in fact, {\it fully invariant}, i.e., is  invariant under any
endomorphism  of $F_r$. This implies that the group $G$ has {\it a
lot} of endomorphisms because any map on the generators of $G$ can
be extended (by the homomorphic property) to an endomorphism of $G$.
Thus, if  $G$ has $r$ generators and $m$ elements altogether, then
it has $m^r$ endomorphisms. Even if $r$ is very small (say, $r=3$),
this number is huge because, as we have just seen, $G$ has at least
$p^{r^2+r}$ elements, so with a 100-bit $p$, we are going to have at
least $2^{3600}$ endomorphisms. Of course, we want our endomorphism
$\phi$ not to have short cycles (i.e., if $\phi^m = \phi^n$, then
$|m-n|$ has to be quite large). This is easier to guarantee if
$\phi$ is actually an automorphism because then we can sample from
automorphisms having a large order, and these correspond to matrices
from $GL_r(\Z_{p^2})$ that have large order. Sampling matrices of
large order from that group is not completely trivial, but we leave
this outside of the scope of this survey. Here we just mention that
for most automorphisms of $G$, relevant security assumptions will
not have a compact form like that in Section \ref{Security} because
a product of the form $\phi^{m-1}(g) \cdots \phi^{2}(g) \cdot
\phi(g) \cdot g$ (see the general protocol in our Section
\ref{protocol_holomorph}) typically does not simplify much.

\end{document}